\documentclass[a4paper,12pt]{amsart}

\usepackage{graphicx}
\usepackage{psfrag}
\usepackage[centering]{geometry}
\usepackage{amssymb}
\usepackage[colorlinks=true]{hyperref}
\usepackage[all]{hypcap}

\title[On some ground state components of the $O(1)$ loop model]{On some ground state components of the $O(1)$ loop model}
\author[T.~Fonseca]{Tiago~Fonseca}
\address{Tiago Fonseca, LPTHE (CNRS, UMR 7589), Univ Pierre et Marie Curie-Paris6, 75252 Paris Cedex, France.}
\email{fonseca\,@\,lpthe.jussieu.fr}
\author[P.~Zinn-Justin]{Paul~Zinn-Justin}
\address{Paul Zinn-Justin, LPTMS (CNRS, UMR 8626), Univ Paris-Sud, 91405 Orsay Cedex, France; and LPTHE (CNRS, UMR 7589), Univ Pierre et Marie Curie-Paris6, 75252 Paris Cedex, France.}
\email{pzinn\,@\,lpthe.jussieu.fr}
\thanks{PZJ was supported by
EU Marie Curie Research Training Networks
``ENRAGE'' MRTN-CT-2004-005616, ``ENIGMA'' MRT-CT-2004-5652,
ESF program ``MISGAM''
and ANR program ``GIMP'' ANR-05-BLAN-0029-01.}

\numberwithin{equation}{section}

\newtheorem*{theo*}{Theorem}
\newtheorem*{conj*}{Conjecture}
\newtheorem{lemma}{Lemma}
\newtheorem*{defi}{Definition}
\newtheorem*{coro*}{Corollary}

\date{\today}

\begin{document}

\begin{abstract}

We address a number of conjectures about the ground state $O(1)$ loop model, computing in particular two infinite series of partial sums of its entries and relating them to the enumeration of plane partitions.  Our main tool is the use of integral formulae for a polynomial solution of the quantum Knizhnik--Zamolodchikov equation.

\end{abstract}

\maketitle
{\footnotesize\tableofcontents}

\section{Introduction}

The present work stems from the investigation of the so-called Razumov--Stroganov (RS) conjecture \cite{RS-conj} (see also \cite{BdGN-XXZ-ASM-PP,dG-review}), which is a surprising connection between the model of Fully Packed Loops (FPL) and the ground state of the $O(1)$ loop model. Although various corollaries and side results were proved as byproducts of attempts to prove the RS conjecture \cite{artic34,artic38}, the conjecture itself remains unproven.

In a series of recent papers \cite{artic41,artic42,artic43,artic44,artic45},
it was shown how integral formulae for a certain polynomial solution of
the quantum Knizhnik--Zamolodchikov equation allowed to produce various explicit results, including a connection
with the enumeration of plane partitions (inspired by the conjecture \cite{DF-qKZ-TSSCPP}).
The same general strategy will be used in
this article in order to perform some computations on the ground state components of $O(1)$ loop model.
We shall address some conjectures of Zuber stated in \cite{Zuber-conj}; 
note that thanks to the Razumov--Stroganov conjecture, they can be considered as either 
conjectures on the FPL model (in which case several of them, including
one to be discussed below, were proved in \cite{CK,CKLN}), or on the $O(1)$ loop model,
the latter point of view being ours.
We shall also obtain some new results connecting
the enumeration of certain classes of Plane Partitions and Non-Intersecting Lattice Paths (NILPs) 
with matchings of the form $(\pi)_p$ and $(_p\alpha$ (see section 2 for an
explanation of the notation), and prove a conjecture of \cite{MNdGB}.
These can be thought of as a small step towards a bijection between
Totally Symmetric Self--Complementary Plane Partitions (TSSCPPs) and Alternating Sign Matrices
(ASMs), the latter being in trivial bijection with FPLs, 
since they provide families of equinumerous classes of TSSCPPs and ASMs.

The paper is organized as follows. In the second section we present the various models involved: $O(\tau)$ loop model, FPLs, NILPs and TSSCPPs.
In the third section we describe the quantum Knizhnik--Zamolodchikov equation and the relevant polynomial solution.
In the last two sections we obtain some properties of the entries of this polynomial solution indexed by matchings of the form $(\pi)_p$ and $(_p\alpha$, respectively. In particular in each case, we describe the corresponding counting problem for NILPs and TSSCPPs.% with these polynomial solutions in the homogeneous limit. For the case $(\pi)_p$ we are able to calculate the limit of high $p$, proving one of Zuber's conjectures. 

\section{The models}

In this section we describe the various models that are relevant to this work:
although the latter is concerned with the $O(1)$ loop model, we mention here
for motivation the Fully Packed Loop (FPL) model and their
connection (the Razumov--Stroganov conjecture \cite{RS-conj}).
We also introduce
Non-Intersecting Lattice Paths and Plane Partitions.

\subsection{$O(\tau)$ loop model}
Loop models are an important class of two-dimensional statistical lattice models; indeed they present a wide range of critical phenomena and many classical models can be mapped into a loop model. Here we consider the $O(\tau)$ loop model.

\begin{figure}
\centering
\includegraphics[scale=0.4]{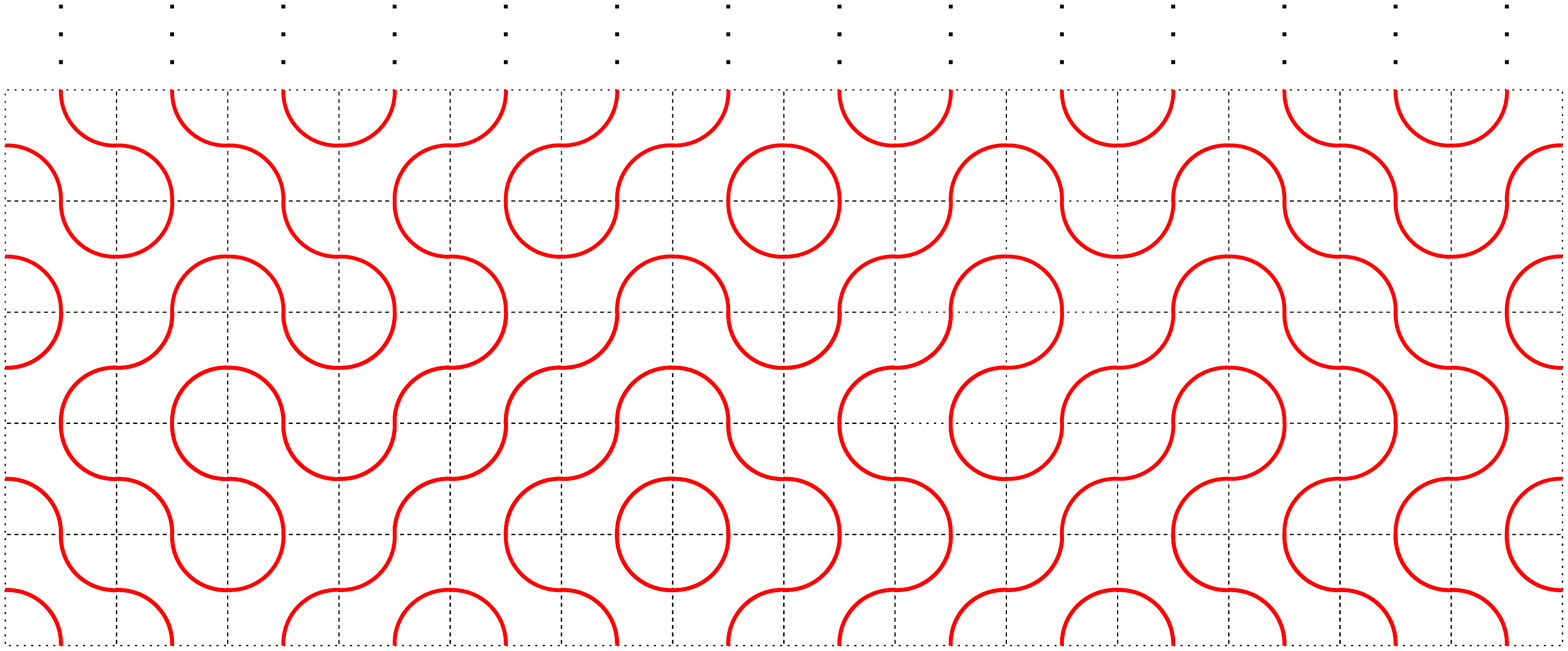}
\caption{\label{O_1} An example of a state in the $O(\tau)$ loop model, here with $n=7$. Each row is filled with plaquettes turning left or right. The left boundary is identified with the right boundary forming a cylinder. At each closed loop is given the weight $\tau$.}
\end{figure}

Consider a semi-infinite cylinder. Each row is made of $2n$ plaquettes which can contain two possible drawings, as on figure \ref{O_1}. We give the weight $\tau$ at each closed loop.

\subsubsection{The space of states}

\begin{figure}
 \centering
 \includegraphics[scale=0.4]{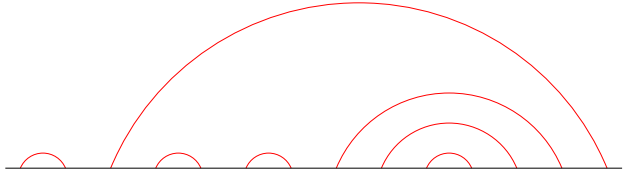}
 \caption{The same example, but here represented in the form of matching (also called link pattern). \label{boucles}}
\end{figure}

In order to set up a transfer matrix approach, we define a state of this model to be the connectivity of the $2n$ points at the bottom. Figure \ref{boucles} represents the connectivity which corresponds to figure \ref{O_1}. The diagrams thus obtained are called matchings (or sometimes, link patterns). The number of such diagrams of size $2n$ is the Catalan number $c_n$:
\[
 c_n=\frac{(2n)!}{n!(n+1)!}
\]

Sometimes it is more convenient to represent states with a series of parentheses. The bijection consists in putting a `$($' at the start of each arch and a `$)$' at the end; thus, our example becomes:
\begin{equation*}
 ()(()()((())))
\end{equation*}

We use the notation $(\pi)_p$ to represent $p$ parentheses surrounding a matching $\pi$: \[(\pi)_p=\underbrace{(\ldots(}_p\pi\underbrace{)\ldots)}_p\]
and $()^p$ for $p$ successive $()$:
\[()^p=\underbrace{()\ldots()}_p\]

\begin{figure}
 \centering
\psfrag{a}[0][0][1][0]{$(0,0)$}
\psfrag{b}[0][0][1][0]{$(2n,0)$}
 \includegraphics[scale=0.5]{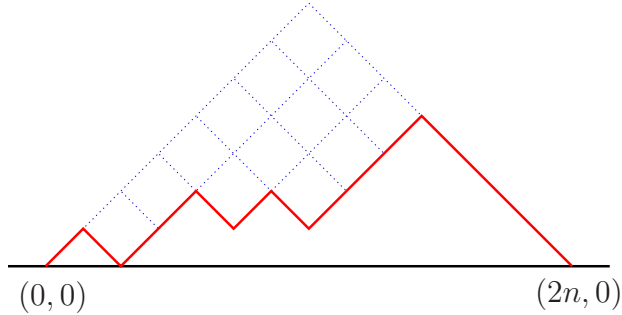}
 \caption{A state of the $O(\tau)$ loop model can be represented by a Dyck path. A Dyck path is composed by $n$ NE steps and $n$ SE steps making a path above the horizontal line (defined by the first and last point). At each `$($' corresponds a NE step and at each `$)$' corresponds a SE step. The blue dotted line represent an Young diagram.\label{dyck}}
\end{figure}

Another way to represent these matchings is using Dyck paths. A Dyck path is a path starting at $(0,0)$ and ending at $(2n,0)$ composed of $n$ NE steps (or $(1,1)$) and $n$ SE steps (or $(1,-1)$), such that the path never goes under the horizontal line defined by the extreme points. To construct a Dyck path from a matching we replace each opening with a NE step and each closing with a SE step as shown on figure \ref{dyck}.

Finally, one can also represent a Dyck path as a Young diagram included in the $(n-1,n-2,\ldots,2,1)$ Young diagram. The Young diagram is obtained from the Dyck path by constructing the complementary of the path. On figure \ref{dyck} we exemplify how to transform a Dyck path into a Young diagram (and vice-versa).

In what follows all operators act
on the vector space of formal linear combination of the matchings
$ \left|\xi\right>=\sum_\pi\xi_{\pi} \left|\pi\right> $.

\subsubsection{The transfer matrix}
\begin{figure}
 \centering
 \psfrag{a}[0][0][1][0]{$\frac{qz_i-q^{-1}t}{qt-q^{-1}z_i}$}
 \psfrag{b}[0][0][1][0]{$\frac{z_i-t}{qt-q^{-1}z_i}$}
 \psfrag{c}[0][0][1][0]{$t$}
 \psfrag{d}[0][0][1][0]{$z_i$}
 \includegraphics[scale=0.6]{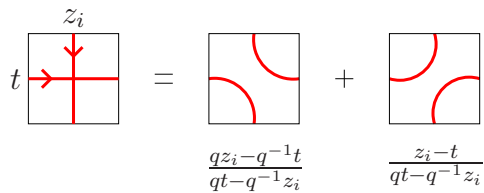}
 \caption{\label{plaquette}The two possible plaquettes and their respective probabilities.}
\end{figure}

The dynamics of the system is encoded in the transfer matrix, which represents one row of plaquettes; each plaquette can have a left turn or a right turn as explained on figure \ref{plaquette}.

\begin{figure}
 \centering
 \includegraphics[scale=0.5]{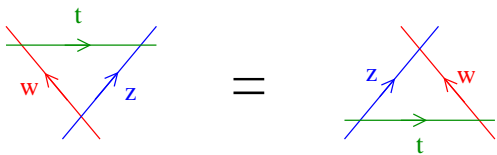}
 \caption{\label{YB-eq}The Yang--Baxter equation.}
\end{figure}

The probability of each plaquette turning left (and in the same way the probability of turning right) is defined by a horizontal parameter $t$ and a vertical parameter $z_i$ (the $i$ indexes the column). The parameters $q$ and $\tau$ are related by $\tau=-q-q^{-1}$. With this parametrization, the weights satisfy the Yang--Baxter equation (figure \ref{YB-eq}).
As a consequence, the transfer matrices satisfy the commutation relation
\[
 [T(t|z_1,\ldots,z_{2n}),T(t'|z_1,\ldots,z_{2n})]=0
\]
so that, assuming diagonalizability of the $T(t|z_1,\ldots,z_{2n})$,
their eigenvectors do not depend on the parameter $t$. We are here especially interested
in the ground state eigenvector, denoted by $\left|\Psi\right>$.

\subsubsection{The Hamiltonian}
We are mostly interested in the homogeneous limit where all the $z_i$ are equal.
In this case one can alternatively define the model using a Hamiltonian.

\begin{figure}
 \centering
 \psfrag{a}[r][0][1.5][0]{$e_5$}
 \includegraphics[scale=0.4]{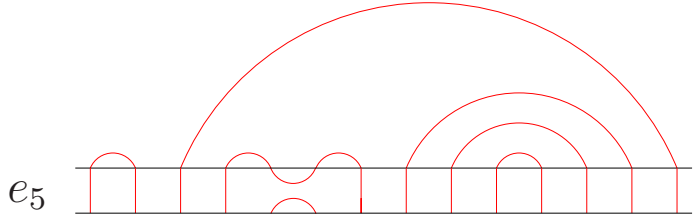}
 \caption{The $e_i$ creates a small arch between $i$ and $i+1$ as exemplified in this figure. If the matching $\pi$ has already a link between $i$ and $i+1$ we get the same state but multiplied by a constant $e_i \pi = \tau \pi$, as the operator $e_i$ leaves $\pi$ invariant forming a closed loop.\label{TL}}
\end{figure}

\begin{defi}
The affine Temperley--Lieb algebra.

Let $e_i$ be the operator that acts as indicated on figure \ref{TL}, the $e_{2n}$ creates an arch between $1$ and $2n$ as our model is defined on a cylinder. This operator obeys the affine Temperley--Lieb algebra:

\begin{align*}
 e_i e_j  &= e_j e_i && \text{if } \left| i-j \ (\textrm{mod }2n)\right| \geq 2\\
 e_i e_{i\pm1} e_i &= e_i  \\
 e_i^2  & = \tau e_i 
\end{align*}
\end{defi}

Consider the Hamiltonian obtained as the logarithmic derivative of the transfer matrix $T(t|1,\ldots,1)$, at $t=1$, which, up to an additive and a multiplicative constant, is given by:
\begin{equation}
 H=-\sum_{i=1}^{2n} e_i \label{hamiltonian}
\end{equation}

\subsubsection{The special case $\tau=1$}
The case $\tau=1$, or equivalently $q=e^{\pm 2 \pi i /3}$, plays a special role.
At this value, the Yang--Baxter equation allows us to write \cite{artic34}:
\begin{equation}
\left( \frac{qz_{i+1}-q^{-1}z_{i}}{qz_{i}-q^{-1}z_{i+1}}I+\frac{z_{i+1}-z_{i}}{qz_{i}-q^{-1}z_{i+1}}e_i \right) \left|\Psi\right> = s_i \left|\Psi\right> \label{YB_e}
\end{equation}
where $s_i$ permute $z_i$ and $z_{i+1}$: $s_i f(z_i,z_{i+1})=f(z_{i+1},z_i)$.

Note that in the case $z_i=1$ for all $i$, the ground state of $H$ (and of $T$) is characterized by:
\[
 H \left|\psi\right> =-2n \left|\psi\right>\\
\]
Here we use the notation $\left|\psi\right>$ for the specialization at $z_i=1$ for all $i$ of $\left|\Psi\right>$.
We decompose our ground state as
$
 \left|\Psi\right>=\sum_\pi \Psi_{\pi} (z_1,\ldots,z_{2n}) \left|\pi\right> 
$
and similarly for $\left|\psi\right>$.

\subsection{Fully packed loops} \label{RS-c}

\begin{figure}
 \centering
 \includegraphics[scale=0.5]{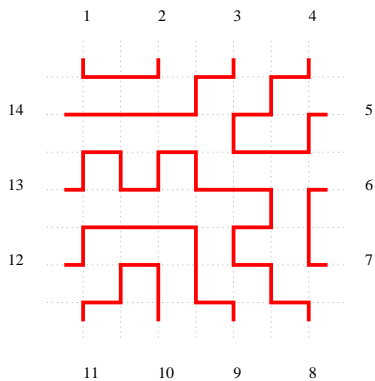}
 \caption{\label{FPM} A fully packed loop configuration is defined on a grid $n\times n$, with paths passing through each vertex. Boundary edges are alternatingly occupied or empty; the numbers label the occupied ones.}
\end{figure}

A fully packed loop (FPL) consists in a grid $n$ by $n$, in which each point has valence two, i.e. each point is connected to two neighbors. We use the boundary conditions defined on figure \ref{FPM} (equivalent to the domain wall
boundary conditions of the six-vertex model).

If we only consider the pairings between the (labelled) exterior points, we can map the configurations of figure \ref{boucles} to matchings. 
Note that this is not an injective function: generally, there are various FPL states that produce the same matching. 
It is therefore natural to consider the number, denoted $A_{\pi}$, of FPL configurations with matching $\pi$. The Razumov-Stroganov conjecture (formulated in \cite{RS-conj}; see also \cite{BdGN-XXZ-ASM-PP,dG-review}) states the following:

\begin{conj*}
Let be $\left|\psi\right>=\sum_\pi \psi_{\pi} \left|\pi\right>$ the ground state of the Hamiltonian 
\[
H=-\sum_{i=1}^{2n} e_i
\]
at $\tau=1$, with the normalization condition $\psi_{()_n}=1$. Then
\begin{equation*}
 A_{\pi}=\psi_{\pi}
\end{equation*}
\end{conj*}

\subsection{Non-Intersecting Lattice Paths}
These paths are defined on a lattice and connect a set of initial points to a set of final points with certain conditions (see Ref. \cite{Lind,GV} for the general framework). The most important feature of Non-Intersecting Lattice Paths (NILPs) is that the various paths do not touch one another, this will be important on the process of counting them using the Lindstr\"om--Gessel--Viennot (LGV) formula \cite{Lind,GV}.

\begin{figure}
 \centering
 \includegraphics[scale=0.7]{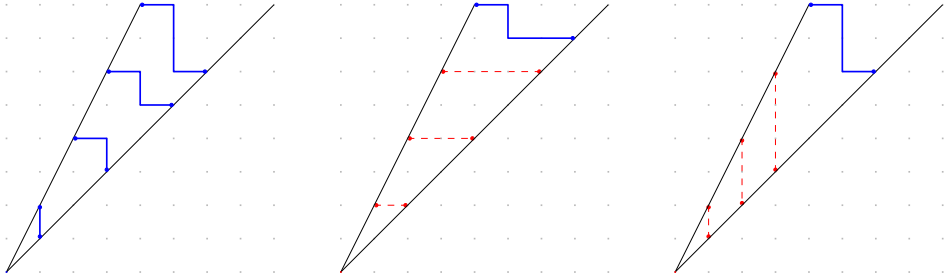}
 \caption{\label{F_G_NILP} In the left we see a NILP composed by 5 paths. In the middle, we see another NILP, but the $p$ dashed red line are fixed, the function that counts these NILPs is called $\mathcal{F}_{p-1,r+1}$ where $n=p+r$ is the number of paths (here $5$), $p$ the fixed paths (here $4$). In the right we find something similar, but the dashed red lines are now vertical and we associate them the function $\mathcal{G}_{p,r}$.}
\end{figure}

We shall be interested in some classes of NILPs. 
We pick the area defined by $x\le y\le 2x$.
We define a path as starting at a point $(i,2i)$ with $i$ a non-negative integer less than $n$.
Each path is composed of steps forward and down, ending in some point in the line $y=x$. We are interested in the set of $n$ paths exemplified on figure \ref{F_G_NILP}. Each vertical step has a weight $\tau$, for example, the first figure in \ref{F_G_NILP} has weight $\tau^4$.

In the sections \ref{NILP_F} and \ref{NILP_G} we will count the number of these NILPs. 
For this it will be important to consider some constraints in the NILPs. 
We can impose the first $p$ paths to be horizontal or to be vertical (we consider the null path at $i=0$). 
The associated counting functions will be called $\mathcal{F}_{p-1,r+1}$ and $\mathcal{G}_{p,r}$, respectively.

\subsection{Totally Symmetric Self--Complementary Plane Partitions}
\begin{figure}
 \centering
 \includegraphics[scale=0.4]{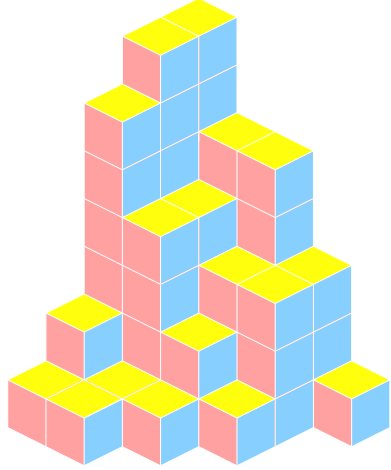}
 \caption{\label{PP}A plane partition.}
\end{figure}

Pictorially, a plane partition is a stack of cubes pushed into a corner, gravity pushing them to the corner, and drawn in a isometric perspective as exemplified on figure \ref{PP}.

Plane partitions were first introduced by MacMahon in 1897. 
In the pictorial representation, Totally Symmetric and Self-Complementary Plane Partitions (TSSCPPs) are Plane Partitions inside a $2n\times 2n\times 2n$ cube which are invariant under the following symmetries:
all permutations of the axes of the cube of size $2n\times 2n\times 2n$; and taking the complement, that is putting cubes where they are absent and vice versa, and flipping the resulting set of cubes to form again a Plane Partition.

As we will see in sections \ref{F_TSSCPP} and \ref{G_TSSCPP} there is a bijection between TSSCPPs and the class of NILPs described before. More precisely, there is a bijection between NILPs with fixed paths and Punctured TSSCPPs.

\section{Polynomial solution of the quantum Knizhnik--Zamolodchikov equation}

At the value $\tau=1$ of the $O(\tau)$ loop model, the ground state eigenvalue takes a
simple form, and the ground state components become
polynomials of the inhomogeneities \cite{artic34}.
We now introduce an equation satisfied by these polynomials.
First we define the invertible operator $\rho$
with relations:

\begin{align*}
 \rho e_i &= e_{i+1} \rho && \text{for all }i\in\{1,\ldots,2n-1\}\\
 \rho^{2n}&=1 
\end{align*}

\begin{figure}
 \centering
 \psfrag{a}[0][0][1.5][0]{$\rho$}
 \includegraphics[scale=0.3]{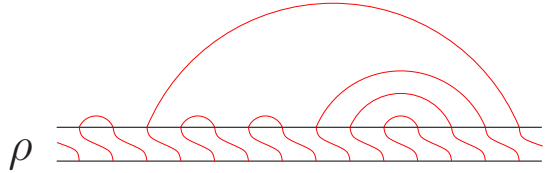}
 \caption{\label{rho}The introduced rotation operator $\rho$ acts in a natural way on the space of matchings.}
\end{figure}

In the loop picture this operator corresponds to the rotation operator, see figure~\ref{rho}.

%Let's consider some polynomials $\Psi_\pi(z_i\ldots,z_{2n})$ indexed by the $c_n$ matchings. We can define the obvious operators:
%\begin{align*}
% \sum_\pi \hat{e}_i \Psi_\pi (z_1,\ldots,z_{2n}) |\pi>&=\sum_\pi \Psi_{\pi} (z_1,\ldots,z_{2n})|e_i \pi>\\
% \sum_\pi \hat{\rho} \Psi_\pi (z_1,\ldots,z_{2n}) |\pi> &=\sum_\pi \Psi_{\pi} (z_1,\ldots,z_{2n})|\rho^{-1}\pi>
%\end{align*}
%where the l.h.s. represents the action on the polynomials and the r.h.s. the action on the matchings.

Consider the following system of equations
for homogeneous polynomials $\Psi_\pi$ of the variables $z_1,\ldots,z_{2n}$ of degree $\delta$:
\begin{itemize}
 \item The \emph{exchange}\/ equation (identical to \eqref{YB_e}):
\begin{equation}
\left( \frac{qz_{i+1}-q^{-1}z_{i}}{qz_{i}-q^{-1}z_{i+1}}I+\frac{z_{i+1}-z_{i}}{qz_{i}-q^{-1}z_{i+1}}e_i \right) \left|\Psi\right> = s_i\left|\Psi\right>
\qquad i=1,\ldots,N-1
 \label{YB_2}
\end{equation}

 \item The \emph{rotation}\/ equation is written in the form:
\begin{equation*}
 \rho^{-1} \Psi_\pi (z_1,z_2,\ldots,z_{2n})=\kappa \Psi_\pi (z_2,\ldots,z_{2n},q^6 z_1)
\end{equation*}
where $\kappa$ is a constant such that $\rho^{-2n}=1$, so that $\kappa^{2n}q^{6\delta}=1$.
\end{itemize}

Although these equations are not what is usually called the quantum Knizhnik-Zamolodchikov (introduced in \cite{FR-qKZ} as a $q$-deformation of the Knizhnik-Zamolodchikov equations), it is easily shown that the solution of this system is also solution to the (level 1) $q$KZ equation.

When $q=e^{2\pi i/3}$ (that is $\tau=1$), we obtain the equations that characterize the ground state of the $O(1)$ loop model. In this case the minimal degree for which there exists a solution is $\delta=n(n-1)$ (all other polynomial solutions will be multiples of this lowest degree solution at $q=e^{2\pi i/3}$).

In \cite{artic43,hdr}, a method is described to construct the solution of this system systematically. 
Schematically, an order is defined on the matchings, and rewrite the \emph{exchange}\/ equation in a triangular form such that it is enough to know $\Psi_\pi$ for the smallest element $\pi=()_n$, in order to be able to compute $\Psi_\pi$, for all $\pi$. In fact this triangular system can be explicitly solved \cite{KL-KL,dGP-factor}.

However this method is not particularly convenient for our purposes, and here we shall use instead an integral formula for $\Psi$ using another basis (presented in section \ref{a_base}).

\subsection{Wheel condition}
Using equation \eqref{YB_2} and knowing that the polynomial has degree $n(n-1)$ we can prove the important relation $\left.\Psi_{\pi}(z_1,\ldots,z_{2n})\right|_{z_k=q^2z_j=q^4z_i}=0$ for all $k>j>i$, called wheel condition (see \cite{Pas-RS}).

The space of homogeneous polynomials in $2n$ variables of total degree $n(n-1)$ which satisfy the wheel condition has been studied in various articles \cite{Kas-wheel,Pas-RS}. The interesting fact is that this space has exactly dimension $c_n$ (the Catalan number and also the number of different matchings of size $2n$), as has been proven in \cite{Pas-RS} and, in a simpler way, in the appendix C of \cite{artic45}.

The proof consists in proving that these polynomials are completely characterized by the values it takes at the $c_n$ points $q^\epsilon:=(q^{\epsilon_1},\ldots,q^{\epsilon_{2n}})$, where $\epsilon_i=\pm 1$ is another way to encode a matching, $-1$ for the opening `$($' and $+1$ for the closing `$)$'.
In what follows,
$\epsilon$ will represent both the matching and the corresponding series of $\pm 1$ depending on the context.

In order to proceed we need the following lemma.

\begin{lemma} \label{recurrencia}
 In a vector $(q^\epsilon)$, we pick one pair $(q^{\epsilon_i},q^{\epsilon_{i+1}})=(q^{-1},q)$. If $\pi$ connects $(i,i+1)$, then we have:
 \begin{equation}
  \Psi_{\pi} (q^{\epsilon})=
    \prod_{j=1}^{i-1}(1-q^3 q^{\epsilon_j}) \prod_{j=i+2}^{2n} (1-q^{-3}q^{\epsilon_j}) \Psi_{\hat{\pi}} (q^{\hat{\epsilon}})\label{pol}
 \end{equation}
where the hat means that we remove the little arch $(i,i+1)$. Otherwise
 \[
  \Psi_\pi (q^\epsilon)=0
 \]
\end{lemma}

For the proof of this lemma see \cite{hdr}. This lemma allows us to calculate the value of $\Psi_\pi(q^{\epsilon})$ but we shall not need the explicit result.

\subsection{Another basis: the $a$-basis\label{a_base}}
Let $a=(a_1,\ldots,a_n)$ be a strictly increasing sequence of integers such that $a_1\geq 1$ and $a_i\leq 2i-1$ for all $i$. We consider the following contour integral:

\begin{multline}
 \Phi_a(z_1,\ldots,z_{2n})=\label{phi_a}
\\(-1)^{\binom{n}{2}}\prod_{1\le i<j\le 2n} (qz_i-q^{-1}z_j) \oint\ldots\oint \prod_{i=1}^n \frac{dw_i}{2\pi i} \frac{\prod_{1\le i<j\le n}(w_j-w_i)(qw_i-q^{-1}w_j)}{\prod_{1\le k\leq a_i}(w_i-z_k)\prod_{a_i<k\le 2n}(qw_i-q^{-1}z_k)}
\end{multline}
where the integral is performed around the $z_i$ but not around $q^{-2}z_i$.

We claim that these polynomials span our vector space of dimension $c_n$.
To prove it we need to check several properties:
\begin{itemize}
 \item Prove that the $\Phi_a$ are indeed, polynomials. It is enough to prove that the expression does not have poles, as done in \cite{artic41}.

 \item Check the total degree ($\delta=n(n-1)$) and the homogeneity. This follows directly from the integral formula, knowing that they are polynomials.

 \item Check the wheel condition. See \cite{artic41} for this proof.

 \item Prove that the $\Phi_a$ are linearly independent. This will be obvious after the calculation of the basis transformation between $\Psi_\pi$ and $\Phi_a$.
\end{itemize}

%\begin{multline}
%\Phi_a(z_1,\ldots,z_{2n})=\\(-1)^{s(\{k\})+\binom{n}{2}}\sum_{\substack{\{k\}=\{k_1,\ldots,k_n\}\\{k_l\neq k_m\ \text{if}\ l\neq m}\\{k_l\leq2l-1}}} \prod_{l<m}(qz_{k_l}-q^{-1}z_{k_m})
%\frac{\displaystyle\prod_{\substack{i<j\\i\notin \{k\}\text{\ or\ }(i=k_l \text{\ and\ } j\leq a_l)}}(qz_i-q^{-1}z_j)}{\displaystyle \prod_j\prod_{\substack{i\leq a_j\\i\notin \{k\}\text{\ or\ } i>k_j}}(z_{k_j}-z_i)} \label{qKZ-frac}
%\end{multline}
%where $(-1)^{s(\{k\})}$ is the sign of the permutation that orders $\{k\}$ in a strictly increasing series.

%If we want to prove that it has no poles at $z_j=z_i$, we need to calculate the limit $z_j\rightarrow z_i$ and show that this is a regular point. We will follow the proof in the article \cite{DFZJ-qKZ-TSSCPP}.

%Two cases arises, or $i$ and $j$ belongs both to $\{k\}$ or only one (let it be $z_i$), otherwise the point is regular.
%For a fixed $\{k\}$ including both $z_i$ and $z_j$, suppose that $i=k_s\leq a_s$, $j=k_r \leq a_r$ and $i<j$, so in order to have a pole we need $i<j=k_r\leq a_r$ and $j\leq a_s$. In this case, there is also a pole in $i=k_r$ and $j=k_s$. It's easy to prove that the two poles are symmetric in the limit, annulling themselves in the limit.
%The second case can be proved in a similar way. 

%The degree follows immediately from simple counting.

%The proof that these polynomials obeys to the wheel condition it's also made in the article \cite{DFZJ-qKZ-TSSCPP}, and therefore we will skip the proof.

\subsection{Basis transformation} 

We assume that the $\Phi_a$ are linearly independent.
We define the basis transformation as follows:
\begin{equation}
 \Psi_{\pi}(z_1,\ldots,z_{2n})=\sum_a \tilde{C}_{\pi,a}\Phi_a(z_1,\ldots,z_{2n})
\end{equation}

And the inverse transformation:
\begin{equation}
 \Phi_{a}(z_1,\ldots,z_{2n})=\sum_{\pi}C_{a,\pi}\Psi_{\pi}(z_1,\ldots,z_{2n})
\end{equation}

Note that the polynomials $\Psi_\pi$ can be multiplied by some constant without changing their characteristics, so we can define:
\[
 \Psi_{()}(z_1,z_2)=\Phi_{1}(z_1,z_2)=1
\]

\subsubsection{Calculating the coefficients $C_{a,\pi}$} \label{coef}

We know that these coefficients are unique and fully determined by the $c_n$ points $(q^{\epsilon})$. Therefore we only need these $c_n$ points. Appendix A of \cite{artic41} describes how to perform this calculation graphically. We shall sketch here the process. For this, we will allow all $a=(a_1,\ldots,a_{n})$ such that $a_i\leq a_{i+1}$ with $1\leq a_i \leq 2i-1$ for all $i$.

We pick a little arch in $\pi$, say between $i$ and $i+1$. 
If there are no $a_j=i$ the $\Phi_a$ is zero. So, $a$ is of the form 
\[
 a=(a_1,\ldots,\underbrace{i,\ldots,i}_k,\ldots,a_n)
\]

After a tedious calculation we obtain:
\begin{equation}
 \Phi_a (q^\epsilon)=\prod_{j=1}^{i-1}(1-q^3 q^{\epsilon_j})\prod_{j=i+2}^{2n}(1-q^{-3}q^{\epsilon_j}) U_{k-1} \Phi_{\hat{a}}(q^{\hat{\epsilon}}) 
\end{equation}
where $\hat{a}=(a_1,\ldots,\underbrace{i-1,\ldots,i-1}_{k-1},\ldots,a_n-2)$ and 
\[U_{k-1}=\frac{q^k-q^{-k}}{q-q^{-1}}\]
is a polynomial of $\tau$ of degree $k$.

Observe that the pre-factor is exactly the same as in \eqref{pol}. So, we get:
\[
 C_{a,\pi}=U_{k-1} C_{\hat{a},\hat{\pi}}
\]

The fact that $C_{1,()}=1$ provides an inductive method of calculation.

There is
to our knowledge no direct method to compute $\tilde{C}_{\pi,a}$ explicitly.
We shall use an Ansatz later for some entries of $\tilde{C}$ 
and confirm it by checking it at all values of the form $(q^\epsilon)$.

\subsubsection{Triangularity of the transformation of basis} \label{triangle}

Now we take some strictly increasing set $a=(a_1,\ldots,a_n)$ such that $a_i\leq 2i-1$ and the set of openings in the matching $\pi$ called $(\pi_1,\ldots,\pi_n)$, in this notation, we see that the two objects are essentially the same. We define a partial order: $\pi\leq a$ if and only if $\pi_i\leq a_i$ for all $i$. We claim that:

\begin{lemma}
 For $C_{a,\pi}$ to be different from zero, $\pi$ must be smaller or equal to $a$. If $\pi_i=a_i$ for all $i$, $C_{a,\pi}=1$.
\end{lemma}

To prove that we shall use the geometric method presented in section \ref{coef}.
We pick an arch in $\pi$, going between $i$ and $j$.
If we use the geometric method of reduction, we see that we need at least $(j-i)/2$ $a_k$ such that $i\leq a_k<j$, applying this in all arches in $\pi$ we see that we need $\pi\leq a$.

If $\pi_i=a_i$ for all $i$, the calculation of $C_{a,\pi}$ is really simple. 

So, if we set a total order which respects the partial order defined before, $C_{a,\pi}$ is a triangular matrix with $1$'s on the diagonal.

The two facts together prove that the transformation is invertible, therefore the $\Phi_a$ are linearly independent.

\subsubsection{Coefficients as polynomials of $\tau$}
The study of these coefficients leads to the following property:
\begin{lemma}\label{boites}
 Let $Y_a$ be the Young diagram $(\lambda_1,\ldots,\lambda_r)$ corresponding to $a=(\ldots,a_i=\lambda_{n-i+1}+i,\ldots)$, and let $Y_\pi$ be the Young diagram corresponding to the matching $\pi$. 

The coefficients $C_{a,\pi}$ are polynomials in $\tau$, more precisely:

\begin{equation}
 C_{a,\pi}=\begin{cases}
            0&\text{if }Y_\pi\nsubseteq Y_a\\
	    1&\text{if }Y_\pi=Y_a\\
	    P_{a,\pi}(\tau)&\text{if }Y_\pi\subset Y_a
           \end{cases}
\end{equation}
where $P_{a,\pi}(\tau)$ is a polynomial of $\tau$ with degree $\delta_{a/\pi}\leq |Y_a|-|Y_\pi|-2$.
\end{lemma}

We leave the proof of this lemma to appendix \ref{skew}.

Observe that this lemma remains true for the coefficients $\tilde{C}_{\pi,a}$, by using triangularity of $C$
and the fact that the diagonal elements are one.

\section{Study of entries of the type $(\pi)_p$}

\begin{figure}

\centering
\psfrag{a}[0][0][1][0]{$p$}
\psfrag{b}[0][0][1][0]{$\pi$}
\includegraphics[scale=0.4]{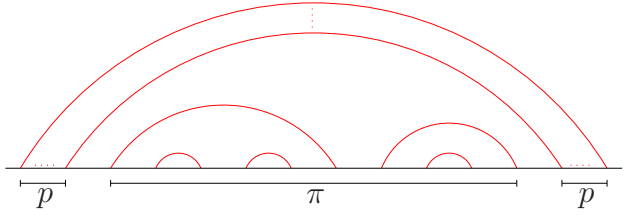}
\caption{A matching $(\pi)_p$ with $p$ arches surrounding a link pattern $\pi$. \label{(p)p}}

\end{figure}

In general, the computation of the polynomials $\Psi_\pi$ is complicated, and there is no general closed formula.
But, there are some exceptions. In this section we study the polynomials indexed by configurations of the type $( \pi )_p$, i.e. a given configuration surrounded $p$ parenthesis (see figure \ref{(p)p} for an example). In subsection 4.3 we present some properties of the polynomials for high $p$.

In all that follows, $\pi$ is a link pattern of
size $2r$, so that $( \pi )_p$ has size $2n$ with $n=r+p$.

\subsection{$a$-Basis}

As done in other articles \cite{artic41} and explained in \ref{coef} we decompose our polynomials in the $a$-basis. In this case, we find that:

\begin{lemma} \label{reduire}
We have the following decomposition
\begin{equation}
\Psi_{(\pi)_p}(z_1,\ldots,z_{2n})=\sum_{\substack{1 \leq a_1<\ldots<a_r < 2r\\a_i\leq 2i-1}} \tilde{C}_{\pi , a} \Phi_{1,\ldots,p,p+a_1,\ldots,p+a_r}(z_1,\ldots,z_{2n}) \label{bas_(p)}
\end{equation}
where the coefficients are the same that occur in
\begin{equation}
\Psi_{\pi}(z_1,\ldots,z_{2r})=\sum_{\substack{1 \leq a_1<\ldots<a_r < 2r\\a_i\leq 2i-1}} \tilde{C}_{\pi , a} \Phi_{a_1,\ldots,a_r}(z_1,\ldots,z_{2r}) \label{bas_p}
\end{equation}
\end{lemma}

This is, if we find the $\tilde{C}_{\pi,a}$ which satisfy the equation \eqref{bas_p}, these same coefficients solve the equation \eqref{bas_(p)}.

One checks that the triangularity forces $\Phi_{1,\ldots,p,p+a_1,\ldots,p+a_r}$ to be written as a linear combination of entries of the type $\Psi_{(\pi)_p}$, and these coefficients do not depend on the value of $p$. When one inverts the transformation only these coefficients will matter, proving the lemma.

\subsection{Reduction to size $r$}

Now, for a given link pattern $\pi$ of size $2r$, 
we can calculate polynomials $\psi_{(\pi)_p}$ for all $p$. For example:

\begin{align}
 \psi_{(()())_p} &= (n-1)\tau \nonumber\\
 \psi_{(()()())_p} &=\frac{n-2}{6}\tau(2\tau^2n^2-5n\tau^2+3\tau^2+6) \label{ex}\\
 \psi_{((())())_p}&=\frac{(n-2)(n-1)}{2}\tau^2 \nonumber
\end{align}
where $\tau=-(q+q^{-1})$.
These are the steps of the calculation:
\begin{itemize}
 \item For given $p$ and $r$ we compute the $C_{a,\pi}$ for all $\pi$ and $a$ of size $r$.

 \item We invert this relation.

 \item We calculate 
\begin{multline*}
 \Phi_{1,\ldots,p,p+a_1,\ldots,p+a_r}(z_1,\ldots,z_{2n})=(-1)^{\binom{n}{2}}\prod_{1\leq i<j\leq p} (qz_i-q^{-1}z_j)\prod_{p < i<j\leq 2n} (qz_i-q^{-1}z_j)\\ \times\oint \ldots \oint \prod_{i=1}^r \frac{dw_i}{2\pi i} \frac{\prod_{1\leq i<j\leq r} (w_j-w_i) (qw_i-q^{-1}w_j) \prod_{1\leq j \leq p}(qz_j-q^{-1}w_i)}{\prod_{p<k\leq a_i+p}(w_i-z_k)\prod_{a_i+p<k\leq 2n}(qw_i-q^{-1}z_k)}
\end{multline*}
where the integration in the first $p$ variables is already performed.

 \item We use the variable transformation 
\begin{equation}
 u_i  =  \frac{w_i - 1}{q w_i -q^{-1}} \label{w_u}
\end{equation}
to calculate the limit where $z_i=1$ for all $i$, and to finally get:
\begin{equation*}
 \phi_{1,\ldots,p,p+a_1,\ldots,p+a_r}=\oint\ldots\oint \prod_{l=1}^{r}\frac{du_l}{2\pi i u_l^{a_l}} \prod_{l<m\leq r} (u_m-u_l)(1+ \tau u_m + u_l u_m)(1+\tau u_m)^p
\end{equation*}

\end{itemize}

\subsection{Expansion for high $p$\label{theoreme}}
Zuber conjectured the polynomial dependence in $p$ and
large $p$ behavior of the number of Fully Packed Loop configurations with connectivity
$(\pi)_p$ (conjecture 6 in \cite{Zuber-conj}). This was subsequently proved in \cite{CKLN}.
Alternatively, due to the Razumov--Stroganov conjecture, 
one can expect the same behavior for the ground state entries of the $O(1)$ loop model.
Here we generalize it to the $q$KZ solution for any $\tau$:
\begin{theo*}
 For matchings of the type $(\pi)_p$, the polynomials $\psi_\pi$ can be written in the following form:
\[
 \psi_{(\pi)_p}=\frac{1}{|Y|!} P_Y(\tau,n)
\]
where $Y$ is the Young diagram defined by $\pi$, $|Y|$ its number of boxes, and $P_Y(\tau,n)$ is a polynomial in $n$ and $\tau$ of degree $|Y|$ in each variable with integer coefficients.

In the limit of high $n$, the polynomials behave like
\[
 \psi_{(\pi)_p}\approx\frac{\dim Y}{|Y|!}(n\tau)^{|Y|}
\]
where $\dim Y$ is the dimension of the irreducible representation of the symmetric group associated to $Y$.
\end{theo*}

As the basis transformation is triangular, we can write: 
\[
\psi_{(\pi)_p}=\phi_{1,\ldots,p,p+a_1,\ldots,p+a_r}+\sum_{b<a} \tilde{C}_{\pi,b} \phi_{1,\ldots,p,p+b_1,\ldots,p+b_r}
\]
where $a$ is equivalent to $\pi$ and $b<a$ means that the Young diagram of $b$ is inside of the Young diagram of $a$. We denote the Young diagram corresponding to $a$ by $Y_a$.

Note that, by lemma \ref{boites}, $\tilde{C}_{\pi,b}$ are polynomials of $\tau$ with integer coefficients and exponent no more than $|Y_a|-|Y_b|-2$.

We now prove that the integral of the first term, corresponding to the largest Young diagram in the decomposition, is a polynomial of $n$ with the asymptotic behavior of the theorem. The other terms
will possess the same polynomiality property, and they will be of lower degree in $n$ and $\tau$
(noting that the power of $\tau$ in the non-diagonal elements is less than $|Y_a|-|Y_b|$ and does not affect our conclusion). 

We want to calculate the integral 
\begin{equation}
 \phi_{1,\ldots,p,p+a_1,\ldots,p+a_r}=\oint\ldots\oint\prod_{i=1}^{r}\frac{du_i}{2\pi i u_i^{a_i}}\prod_{i < j\leq r} (u_j-u_i)(1+\tau u_j + u_j u_i )(1+\tau u_j)^p
\end{equation}

We replace the term $\prod_{i<j}(1+\tau u_j +u_i u_j)$ with $\prod_{i<j}(1+\tau u_j)=\prod_i (1+\tau u_i)^{i-1}$, because any term with $u_i u_j$ in the product is formally identical to the contribution of a smaller diagram.
We then compute
\begin{align*}
 \oint\ldots\oint \prod_{i=1}^r\frac{du_i}{2\pi i u_i^{a_i}} (1+\tau u_i)^{p+i-1} &\prod_{j>i} (u_j-u_i)\\
 &=\sum_\sigma (-1)^\sigma \oint\ldots\oint \prod_{i=1}^r\frac{du_i}{2\pi i u_i^{a_i}}(1+\tau u_i)^{p+i-1} u_i^{\sigma_i-1}\\
 &=\sum_\sigma (-1)^\sigma \oint\ldots\oint \prod_{i=1}^r\frac{du_i}{2\pi i u_i^{i+\lambda_i+1-\sigma_i}}(1+\tau u_i)^{p+i-1}\\
 &=\tau^{|Y|}\sum_\sigma (-1)^\sigma \prod_{i=1}^r\binom{n-r+i-1}{i+\lambda_{n-i+1}-\sigma_i}
\end{align*}
where $\lambda_{n-i+1}=a_i-i$ is the size of each row in the Young diagram and $(-1)^\sigma$ is the sign of the permutation $\sigma$.
We obtain that the coefficients can be written as a sum of integers divided by $\prod_i (i+\lambda_i-\sigma_i)!$ which divide $|Y|!$ as a consequence of $\sum_i(i+\lambda_i-\sigma_i)=|Y|$. 

The dominant contribution as a function of $n$ is 
\begin{align*}
 \tau^{|Y|}\sum_\sigma (-1)^\sigma \prod_{i=1}^r \frac{n^{i+\lambda_{n-i+1}-\sigma_i}}{(i+\lambda_{n-i+1}-\sigma_i)!}
&=(\tau n)^{|Y|} \sum_\sigma (-1)^\sigma \prod_{i=1}^r \frac{1}{(i+\lambda_{n-i+1}-\sigma_i)!}\\
&=(\tau n)^{|Y|} \frac{\dim Y}{|Y|!}
\end{align*}
where the last equality can be found, for example, in the fourth chapter of \cite{FH-book}.

\subsection{Sum rule}
Pick an $a$ of the form $(a_1,\ldots,a_r)$, with $a_i=2i-1$ or $a_i=2i-2$.\footnote{Observe that if $a_1=0$, $\Phi_a=0$.} Let $\mathcal{L}(a)$ be the set of matchings whose openings on odd sites are exactly the odd elements in $a$.

From \cite{artic41}, section 3.3, we know:
\[
 \Phi_a=\sum_{\pi\in\mathcal{L}(a)} \Psi_\pi
\]

It follows that, using lemma \ref{reduire}:
\begin{equation}
  \sum_{\substack{a=(1,\ldots,p,p+a_1,\ldots,p+a_r)\\a_i=2i-1\text{ or }2i-2}}\tau^{r^2-\sum_i a_i}\Phi_{a}=\sum_{\pi} 
 \tau^{o_\pi} \Psi_{(\pi)_p} \label{w_sum}
\end{equation}
where $o_\pi$ counts the openings in even sites, of the matching $\pi$, and $\left(r^2-\sum_i a_i\right)$ counts the number of even $a_i$ in $a$, as $o_\pi=r^2-\sum_i a_i$ if $\pi\in\mathcal{L}(a)$.

In \eqref{w_sum}, at $z_i=1$ for all $i$, the l.h.s.\ is equal to:
\[
 \oint\ldots\oint \prod_{l=1}^r\frac{du_l}{2\pi i u_l^{2l-1}} \prod_{l<m\leq r} (u_m-u_l)(1+\tau u_m + u_l u_m)(1+\tau u_m)^{p+1}
\]
but this is exactly 
\[
 \phi_{1,\ldots,p+1,p+2,p+4,\ldots,p+2r}=\psi_{(()^r)_{p+1}}
\] 

We can now state the main result of this section:
\begin{theo*}
 Let $o_\pi$ count the number of arches of $\pi$ opening at an even site, we have the following result:
\begin{equation}
 \sum_{\pi\text{ of size }2r} 
 \tau^{o_\pi} \psi_{(\pi)_p} = \psi_{(()^r)_{p+1}}
\end{equation}
\end{theo*}

At $\tau=1$, we get the proof of conjecture 8.i of \cite{Zuber-conj}, but for the $O(1)$ loop model:
\begin{equation}
 \sum_{\pi\text{ of size }2r} 
\psi_{(\pi)_p}=\psi_{(()^r)_{p+1}}
\end{equation}

At $p=0$, we can use the rotation symmetry and obtain the already known formula: 
\[\psi_{()^{r+1}}= \sum_{\pi\text{ of size }2r} 
\psi_{\pi}\] 

\subsection{A NILP formula} \label{NILP_F}
\begin{figure}
 \centering
 \psfrag{a}[0][0][.8][0]{$1$}
 \psfrag{b}[0][0][.8][0]{$2$}
 \psfrag{c}[0][0][.8][0]{$3$}
 \psfrag{d}[0][0][.8][0]{$r$}
 \psfrag{e}[0][0][.8][0]{$1$}
 \psfrag{f}[0][0][.8][0]{$2$}
 \psfrag{g}[0][0][.8][0]{$3$}
 \psfrag{h}[0][0][.8][0]{$L_r$}
 \includegraphics[scale=0.5]{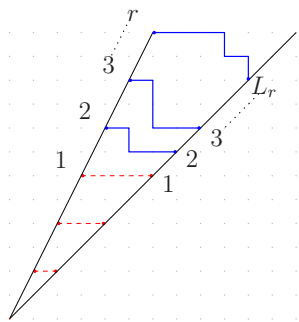}
 \caption{\label{F_NILP} In order to apply the LGV formula we label the starting points of the paths by $1,\ldots,r$ and the ending points by $L_1=1,\ldots,L_r$. As the paths do not intersect each others $L_{i+1}>L_i$.}
\end{figure}

We can interpret the result of the previous subsection in terms of the NILPs previously defined, cf
figure \ref{F_G_NILP}.
We fix $p$ paths as shown in \ref{F_NILP}, and using the LGV formula, we are able to calculate the number of different NILPs. We label the paths with $i=1,\ldots,r$ and call the final locations $L_i$. 

If we only consider one path going from $i$ to $L_i$:
\[
 \mathcal{P}_{i,L_i}=\tau^{2i-L_i-1} \binom{p+i-1}{2 i-L_i-1}
\]
we require that $L_1=1$. We give a weight $\tau$ to each vertical step.

The LGV formula tell us that the number of paths is equal to 
\[
 \mathcal{F}_{p,r}=\sum_{1=L_1<\ldots<L_r} \det \left[\tau^{2i-L_j-1} \binom{p+i-1}{2i-L_j-1}\right]_{1\leq i,j\leq r}
\]

We can use a contour integral form:
\[
 \mathcal{P}_{i,L_i}=\oint \frac{du}{2\pi i} \frac{(1+\tau u)^{p+i-1}}{u^{2i-L_i}}
\]
so, the number of NILPs can expressed by:
\begin{align*}
 \mathcal{F}_{p,r}&= \oint\ldots\oint \prod_{i=1}^r\frac{du_i}{2 \pi i} \sum_{1=L_1<\ldots<L_r} \det \left[ \frac{(1+\tau u_k)^{p+k-1}}{u_k^{2k-L_j}} \right]_{1\leq k,j \leq r}\\
&= \oint\ldots\oint \prod_{i=1}^r\frac{du_i}{2 \pi i} \frac{(1+\tau u_i)^{p+i-1}}{u_i^{2i-1}}\sum_{1=L_1<\ldots<L_r} \det \left[ u_k^{L_j-1} \right]_{1\leq k,j \leq r}\\
&= \oint\ldots\oint \prod_{i=1}^r\frac{du_i}{2 \pi i} \frac{(1+\tau u_i)^{p+i-1}}{u_i^{2i-1}} \frac{1}{1-u_i} \prod_{i<j} \frac{u_j-u_i}{1-u_i u_j} \\
&= \oint\ldots\oint \prod_{i=1}^r\frac{du_i}{2 \pi i} \frac{(1+\tau u_i)^{p+i-1}(1+u_i)}{u_i^{2i-1}} \frac{\prod_{i<j} (u_j-u_i)}{\prod_{i\leq j}(1-u_i u_j)} \\
\end{align*}
where the equality between the second and the third line can be found, for example, in the fourth chapter of \cite{Bressoud}.

Applying the formula obtained in appendix D of \cite{artic45}, we transform the formula into:
\begin{equation}
 \mathcal{F}_{p,r} = \oint\ldots\oint \prod_{i=1}^r\frac{du_i}{2\pi i u_i^{2i-1}} (1+\tau u_i)^p (1+u_i) \prod_{i<j} (u_j-u_i)(1+\tau u_j +u_i u_j)
\end{equation}

We recognize this integral formula:
\[
 \mathcal{F}_{p,r} = 
 \sum_{\pi\text{ of size }2r} 
 \tau^{o_\pi} \psi_{(\pi)_p} = \psi_{(()^r)_{p+1}}
\]

There does not seem to be a simple closed formula for $\mathcal{F}_{p,r}$ even at $\tau=1$, 
as suggested by the expressions of \cite{Zuber-conj} for small values
of $p$.

\subsection{Punctured--TSSCPPs\label{F_TSSCPP}}
\begin{figure}
 \centering
 \includegraphics[scale=0.3]{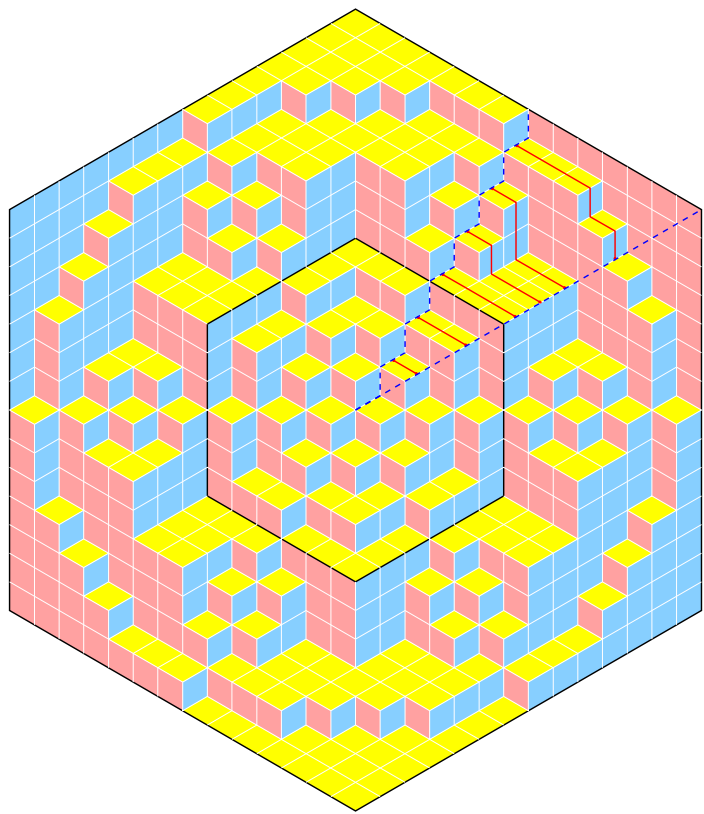}
 \caption{\label{F_PP}This TSSCPP with fixed central hexagon, of size $2p$, corresponds to the NILPs of figure \ref{F_NILP} with $p+1$ fixed horizontal paths. The corresponding NILPs are drawn in the fundamental domain
(the triangle between the two blue dotted lines).}
\end{figure}

Recall that there is a bijection between NILPs and TSSCPPs. As is seen on figure \ref{F_PP}, fixing the first $p+1$ paths to be horizontal amounts to fixing the central hexagon of size $2p$. Observe in addition that the triangle which on the figure contains the paths is a fundamental domain i.e.\ defines the whole TSSCPP. Consequently, $\mathcal{F}_{p,r}$ also counts the number of TSSCPPs with weight $\tau$ for each blue face (the faces containing a vertical step of the paths) in the fundamental domain. See \cite{artic44} for a similar interpretation
of partial sums in a related model in terms of punctured plane partitions.

\section{Study of entries of the type $(_p\alpha$}

\begin{figure}
\centering
\psfrag{a}[0][0][1][0]{$p$}
\psfrag{b}[0][0][1][0]{$\alpha$}
\includegraphics[scale=0.4]{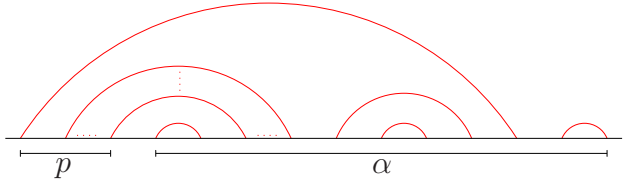}
\caption{A state $(_p \alpha$ is a state that has at least $p$ openings at the left. \label{(p}}
\end{figure}

In this section we consider entries $\psi_\pi$ of the polynomial solution of $q$KZ
(in the homogeneous limit) which correspond to the matchings with $p$ openings at the beginning. Explicitly, such a matching is of the form $(_p \alpha$, where $\alpha$ is not a state but a sequence which contains $n$ closings and $r=n-p$ openings, see figure \ref{(p} for an example. Equivalently these are the matchings for which the $p$ first points are not connected to each other.

\subsection{Basis transformation}
We are here interested in the sum of such polynomials $\psi_{(_p\alpha}$.
Recall that there is no systematic way to obtain the corresponding expression in the $a$-basis.
However, based on some numerical experimentation, the following formula can be guessed:
\begin{lemma}
The sum of the polynomial of the type $\Psi_{(_p\alpha}$ is:
\begin{equation}
 \sum_\alpha \Psi_{(_p\alpha}=\sum_{\substack{a_{i+1}>a_i\\p<a_i\le 2p+2i-1\\a_{i+1}\neq a_i+1\, \text{for all }\,a_i\,\text{even}}} \Phi_{1,2,\ldots,p,a_1,\ldots,a_r} \label{a_G}
\end{equation}

\end{lemma}
The proof consists in evaluating this equality at all $c_n$ points of the type $(q^{\epsilon})$.

It is obvious that the l.h.s.\ evaluated at $(q^\epsilon)$ such that $\epsilon$ can not be written as $(_p\alpha$ is zero.

As to the r.h.s., we integrate the first $p$ variables:
\begin{align*}
 \Phi_{1,\ldots,p,a_1,\ldots,a_n}=& (-1)^{\binom{n}{2}}\prod_{i<j\leq 2n}(qz_i-q^{-1}z_j)\oint\ldots\oint\prod^n_{i=1}\frac{dw_i}{2\pi i}\frac{\prod_{j>i}(w_j-w_i)(qw_i-q{-1}w_j)}{\prod_{b_j\geq i}(w_j-z_i)\prod_{i>b_j}(qw_j-q^{-1}z_i)}\\
=& (-1)^{\binom{n}{2}}\prod_{i<j\leq p} (qz_i-q^{-1}z_j) \prod_{p<i<j\leq 2n}(qz_i-q^{-1}z_j)\\
&\times\oint\ldots\oint\prod_{i=p+1}^n\frac{dw_i}{2 \pi i} \frac{\prod_{j>i}(w_j-w_i)(qw_i-q^{-1}w_j)\prod_{j,i}(qz_i-q^{-1}w_j)}{\prod_{a_j\geq i > p}(w_j-z_i)\prod_{i>a_j}(qw_j-q^{-1}z_i)}
\label{psi_a}
\end{align*}
where $\{b_1,\ldots,b_p,b_{p+1},\ldots,b_n\}=\{1,\ldots,p,a_1,\ldots,a_n\}$.
This sum is zero for all $(q^{\epsilon})$ such that $\epsilon$ do not have $p$ openings at the left. 

We now proceed by induction on $r$.
If $r=n-p=0$ we find that both sides are:
\begin{equation}
 (-1)^{\binom{n}{2}}(q-q^{-1})^{n(n-1)}
\end{equation}

We now want to show that for all $\epsilon$ of the type $(_p \alpha$ the r.h.s.\ satisfies the same recurrence 
as the l.h.s.\
Take an $\epsilon$ which has a pairing (``little arch'') $(i,i+1)$. Using lemma \ref{recurrencia}, we rewrite the l.h.s.\ as:
\begin{equation}
 \sum_\alpha \Psi_{(_p\alpha} (q^{\epsilon})=q^{-(n-1)}\prod_{j=1}^{i-1} (q^{-1}-q^2 q^{\epsilon_j}) \prod_{j=i+2}^{2n} (q^2-q^{-1}q^{\epsilon_j}) \sum_{\hat{\alpha}} \Psi_{(_p\hat{\alpha}} (q^{\hat{\epsilon}})
\end{equation}
where the hat means that we remove the little arch from $(_p\alpha$\footnote{if $\alpha$ does not have a little arch $(i,i+1)$, the term is zero.} and $\epsilon$. It is obvious that $(_p\hat{\alpha}$ passes by all matchings with $n-1$ arches and $p$ openings at the beginning.

We proceed similarly with the r.h.s. We pick the same vector $(q^\epsilon)$. 
We suppose that $i$ is even (for $i$ odd the reasoning is the same).

If $i\notin \{a_j\}$, the expression vanishes. So we pick $a_j=i$, and as $i$ is even we have $a_{j+1}>i+1$.
When we integrate on $w_j$ in expression \eqref{phi_a}, by the rules defined in \ref{coef}, we obtain a similar formula with a reduced vector $\epsilon$ of size $n-1$ that is obtained by removal of the little arch.

The integral formula is modified in the following manner:
\begin{multline*}
\sum_{\substack{a_{i+1}>a_i\\p<a_i\le 2p+2i-1\\a_{i+1}\neq a_i+1\, \text{for all}\,a_i\,\text{even}}} \Phi_{1,2,\ldots,p,a_1,\ldots,a_r}(q^\epsilon)=
q^{-(n-1)}\prod_{j=1}^{i-1} (q^{-1}-q^2 q^{\epsilon_j}) \prod_{j=i+2}^{2n} (q^2-q^{-1}q^{\epsilon_j})
\\[-0.5cm]
\times\sum_{\substack{\hat{a}_{i+1}>\hat{a}_i\\p<\hat{a}_i\le 2p+2i-1\\\hat{a}_{i+1}\neq \hat{a}_i+1\, \text{for all}\,\hat{a}_i\,\text{even}}} \Phi_{1,2,\ldots,p,\hat{a}_1,\ldots,\hat{a}_{r-1}}(q^{\hat{\epsilon}})
\end{multline*}
where 
\begin{equation}
 \hat{a}_i=\begin{cases}
            a_i&i<j\\
	    a_{i+1}-2&\text{otherwise}
           \end{cases}
\end{equation}

This proves the lemma.\qed

Now we can calculate the limit $z_i=1$ for all $i$. Using the change of variables
\[
 u_i  =  \frac{w_i - 1}{q w_i -q^{-1}}
\]
and integrating the first $p$ variables, we obtain:
\begin{multline*}
\sum_\alpha \psi_{(_p\alpha}=
 \sum_{\substack{a_{i+1}>a_i\\p<a_i\leq 2i-1\\a_{i+1}\neq a_i+1\, \text{for all}\,a_i\,\text{even}}}
\oint\ldots\oint \prod_{m=1}^r\frac{du_m}{2\pi i}\frac{(1+ \tau u_m)^p}
{u_m^{a_m-p}}
\\[-0.6cm]
\prod_{1\le l<m\leq r}(u_m-u_l)(1+ \tau u_m +u_m u_l)
\end{multline*} 

To sum over all possible $a$, we can consider only the odd $a_i$ and multiply by $(1+u_i)$, simplifying in this way the conditions:
\begin{multline*}
\sum_\alpha \psi_{(_p\alpha}=
 \sum_{\substack{a_{i+1}>a_i\\p<a_i\leq 2i-1\\a_{i}\text{ odd}}}
\oint\ldots\oint \prod_{m=1}^r\frac{du_m}{2\pi i}\frac{(1+ \tau u_m)^p(1+u_m)}
{u_m^{a_m-p}}
\\[-0.6cm]
\prod_{1\le l<m\leq r}(u_m-u_l)(1+ \tau u_m +u_m u_l)
\end{multline*}

We write $b_i=p-(a_i+1)/2+i$:
\begin{multline*}
\sum_\alpha \psi_{(_p\alpha}=
 \sum_{0\leq b_{i+1}\leq b_i}
\oint\ldots\oint \prod_{m=1}^r\frac{du_m}{2\pi i}\frac{(1+ \tau u_m)^p(1+u_m)}
{u_m^{p+2m-1}}u_m^{2b_m}
\\[-0.2cm]
\prod_{1\le l<m\leq r}(u_m-u_l)(1+ \tau u_m +u_m u_l)
\end{multline*}
The upper bound on $b_1$ was relaxed, since it only excludes zero terms.

A standard calculation gives us the formula:
\[
 \sum_{0\leq b_{i+1}\leq b_i}\prod_i u_i^{2b_i}=\frac{1}{\prod_{m=1}^r(1-\prod_{i=1}^m u_i^2)}
\]

Replacing, we get:
\begin{multline}
\sum_\alpha \psi_{(_p\alpha}=
 \sum_{0\leq b_{i+1}\leq b_i}
\oint\ldots\oint \prod_{m=1}^r\frac{du_m}{2\pi i}\frac{(1+ \tau u_m)^p(1+u_m)}
{u_m^{p+2m-1}(1-\prod_{i=1}^m u_i^{2})}
\\[-0.2cm]
\prod_{1\le l<m\leq r}(u_m-u_l)(1+ \tau u_m +u_m u_l)
 \label{G_loop}
\end{multline}

\subsection{NILPs} \label{NILP_G}
Instead of fixing the first $p$ paths to be horizontal, as in section \ref{NILP_F}, we fix them to be vertical. We count the number of paths with a weight $\tau$ for each vertical step (the first $p$ fixed paths excluded).

\begin{figure}
 \centering
 \psfrag{a}[0][0][.8][0]{$1$}
 \psfrag{b}[0][0][.8][0]{$2$}
 \psfrag{c}[0][0][.8][0]{$3$}
 \psfrag{d}[0][0][.8][0]{$r$}
 \psfrag{e}[0][0][.8][0]{$2$}
 \psfrag{f}[0][0][.8][0]{$5$}
 \psfrag{g}[0][0][.8][0]{$6$}
 \psfrag{h}[0][0][.8][0]{$L_r$}
 \includegraphics[scale=0.5]{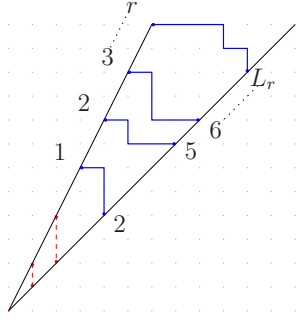}
 \caption{\label{G_NILP} To apply the LGV formula we need to label the starting points of the paths by $1,\ldots,r$ and the ending points by $1 \leq L_1,\ldots,L_r$. As they do not intersect each others $L_{i+1}>L_i$.}
\end{figure}

We use the same method that was used for the $\Psi_{(\pi)_p}$ polynomials. On figure \ref{G_NILP} we show how to label the entries. 

Consider all the paths going from $i$ to $L_i$, we get:
\[
 \mathcal{P}_{i,L_i}=\tau^{p+2i-L_i-1}\binom{p+i-1}{p+2i-L_i-1}
\]

Or, in a contour integral form:
\[
 \mathcal{P}_{i,L_i}=\oint \frac{du}{2\pi i}\frac{(1+\tau u)^{p+i-1}}{u^{p+2i-L_i}}
\]

We apply the LGV formula:
\begin{align*}
 \mathcal{G}_{p,r}&= \oint\ldots\oint \prod_{i=1}^r \frac{du_i}{2 \pi i} \sum_{1=L_1<\ldots<L_r} \det \left[ \frac{(1+\tau u_k)^{p+k-1}}{u_k^{p+2k-L_j}} \right]_{1\leq k,j \leq r}\\
&= \oint\ldots\oint \prod_{i=1}^r\frac{du_i}{2 \pi i} \frac{(1+\tau u_i)^{p+i-1}}{u_i^{p+2i-1}}\sum_{1=L_1<\ldots<L_r} \det \left[ u_k^{L_j-1} \right]_{1\leq k,j \leq r}\\
&= \oint\ldots\oint \prod_{i=1}^r\frac{du_i}{2 \pi i} \frac{(1+\tau u_i)^{p+i-1}}{u_i^{p+2i-1}} \frac{1}{1-u_i} \prod_{i<j} \frac{u_j-u_i}{1-u_i u_j} \\
&= \oint\ldots\oint \prod_{i=1}^r\frac{du_i}{2 \pi i}  \frac{(1+\tau u_i)^{p+i-1}(1+u_i)}{u_i^{p+2i-1}} \frac{\prod_{i<j} (u_j-u_i)}{\prod_{i\leq j}(1-u_i u_j)} \\
\end{align*}

We now use the following identity (similar to the one formulated in \cite{artic41} and
proved in \cite{Zeil-qKZ}), proved in appendix \ref{zeil}:
\begin{multline*}
 \oint\ldots\oint\prod_{i=1}^r \frac{du_i}{2\pi i} \frac{\prod_{i<j}(u_j-u_i)(1+\tau u_j +u_i u_j)}{u_i^{2i+p-1}(1-\prod_{j=1}^i u_j^2)}\Omega(u_1,\ldots,u_r)=\\
=\oint\ldots\oint\prod_{i=1}^r \frac{du_i}{2\pi i} \frac{(1+\tau u_i)^{i-1}\prod_{j>i}(u_j-u_i)}{u_i^{2i+p-1}\prod_{j\geq i}(1-u_j u_i)}\Omega(u_1,\ldots,u_r)
\end{multline*}
where $\Omega(u_1,\ldots,u_r)$ is some symmetric function, here $\Omega(u_1,\ldots,u_r)=\prod_{i=1}^r(1+\tau u_i)^{p}(1+u_i)$, without poles in the integration region. We thus obtain exactly equation \eqref{G_loop}.

In \cite{Kratten-det-TSSCPP},
Krattenthaler gave an explicit formula for $\mathcal{G}_{p,r}$ at $\tau=1$:
\begin{equation}\label{krattenform}
 \mathcal{G}_{p,r}=
 \begin{cases}
  \prod_{i=0}^{r-1}\frac{(3p+3i+1)!}{(3p+2i+1)!(p+2i)!}\prod_{i=0}^{(r-2)/2}(2p+2i+1)!(2i)! &  
   \text{if }r\text{ is even} \\
  2^p\prod_{i=1}^{r-1} \frac{(3p+3i+1)!}{(3p+2i+1)!(p+2i)!} \prod_{i=1}^{(r-1)/2} (2p+2i)!(2i-1)! & 
   \text{if }r\text{ is odd}
 \end{cases}
\end{equation}
Remarkably, these formulae coincide with those conjectured in
\cite{MNdGB} for $\sum_\alpha \psi_{(_p\alpha}$ at $\tau=1$ (more precisely, what was conjectured,
cf their Eqs.~(40--42), 
was the probability that $p$ consecutive points are disconnected from each other
in the $O(1)$ loop model, that is the ratio of $\sum_\alpha \psi_{(_p\alpha)}$ by the full sum).
Correcting a misprint in their Eq.~(42) we have
\begin{gather}\label{nienform}
\mathcal{G}_{p,r}=\frac{S(2(p+r),p)}{S(2p,p)}\\
S(L,p)=\begin{cases}
\frac{\prod_{\ell=1}^{p/2}\prod_{k=\ell}^{2\ell-1}(L^2-4k^2)}{\prod_{\ell=0}^{p/2-1}(L^2-(2\ell+1)^2)^{p/2-\ell}}&\text{$p$ even}\\
\frac{\prod_{\ell=1}^{(p+1)/2}\prod_{k=\ell}^{2\ell-2}(L^2-4k^2)}{\prod_{\ell=0}^{(p-3)/2}(L^2-(2\ell+1)^2)^{(p-1)/2-\ell}}&\text{$p$ odd}
\end{cases}
\end{gather}
The equality of \eqref{krattenform} and \eqref{nienform} can be obtained by direct computation,
treating separately the parities of $p$ and of $r$.

\subsection{Punctured--TSSCPPs\label{G_TSSCPP}}
\begin{figure}
 \centering
 \includegraphics[scale=0.3]{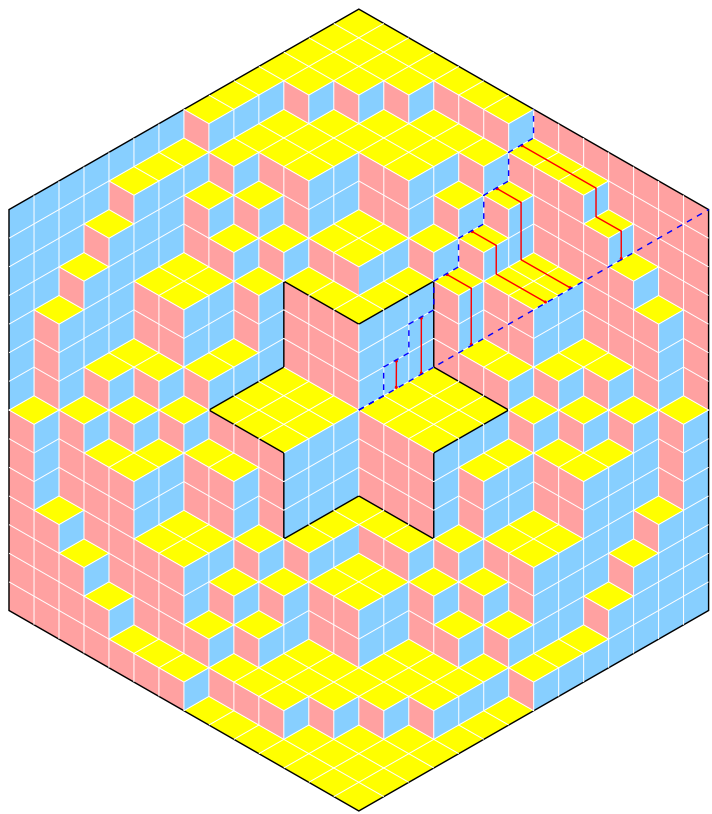}
 \caption{\label{G_PP}This TSSCPP with fixed central hexagonal star, of size $p$, corresponds to the NILP in figure \ref{G_NILP} with $p$ fixed vertical paths. 
%To better understand the bijection we have drawn the tilted NILP, in the figure.
}
\end{figure}

As is seen on figure \ref{G_PP}, fixing the first $p$ paths amounts to fixing a central hexagonal star of size $p$. Consequently, $\mathcal{G}_{p,r}$ also counts the number of TSSCPPs with weight $\tau$ for each blue face in the fundamental domain (the triangle outside the hexagonal star, which on the figure contains the paths).

\appendix

\section{Numerical data}
In this section we give some results, that confirm the theorem in \ref{theoreme}, showing explicitly the form of the polynomials. We use $\tau=1$ for simplicity.\footnote{see \eqref{ex} for some examples with general $\tau$.}

For $r=2$:
\begin{align*}
 \psi_{(()())_p}=&(n-1)
\end{align*}

For $r=3$:
\begin{align*}
 \psi_{(()()())_p}=& \frac{(n-2)}{6}(2n^2-5n+9)\\
 \psi_{((())())_p}=& \frac{(n-2)(n-1)}{2}
\end{align*}

For $r=4$:
\allowdisplaybreaks[1]%allows to break equations if really necessary
\begin{align*}
 \psi_{(()()()())_p}=&\frac{(n-3)(n-1)}{180}(4n^3-32n^3+155n^2-394n+540)\\
 \psi_{(((()))())_p}=&\frac{(n-3)(n-2)(n-1)}{6}\\
 \psi_{((()())())_p}=&\frac{(n-3)(n-2)}{24}(3n^2-7n+16)\\
 \psi_{((())(()))_p}=&\frac{(n-3)(n-2)^2(n-1)}{12}\\
 \psi_{((())()())_p}=&\frac{(n-3)(n-2)(n-1)}{24}(n^2-4n+12)\\
 \psi_{(()(())())_p}=&\frac{(n-3)}{20}(n^4-7n^3+24n^2-48n+60)
\end{align*}

%For $r=5$:

%\begin{align*}
% \psi_{(()()()()())_p}=&\frac{(n-4)(n-2)}{75600}(16n^8-304n^7+3208n^6-21880n^5+103729n^4\\&-339211n^3+746067n^2-1001925 n+737100)\\
% \psi_{((((())))())_p}=&\frac{(n-4)(n-3)(n-2)(n-1)}{24}\\
%\end{align*}

%\begin{align*}
% \psi_{(((()()))())_p}=&\frac{(n-4)(n-3)(n-2)}{120}(4n^2-9n+25)\\
% \psi_{(((())())())_p}=&\frac{(n-4)(n-3)(n-2)(n-1)}{240}(3n^2-11n+40)\\
% \psi_{(()(()())())_p}=&\frac{(n-4)(n-1)}{20160}(45n^6-635n^5+4639m^4-21865n^3+68924n^2\\&-136740n+146160)\\
% \psi_{(((()))(()))_p}=&\frac{(n-4)(n-3)^2(n-2)^2(n-1)}{144}\\
% \psi_{(((()))()())_p}=&\frac{(n-4)(n-3)(n-2)^2(n-1)}{720}(2n^2-11n+45)\\
% \psi_{((()(()))())_p}=&\frac{(n-4)(n-3)}{360}(5n^4-34n^3+127n^2-278n+420)\\
% \psi_{((()()())())_p}=&\frac{(n-4)(n-3)(n-1)}{720}(5n^4-38n^3+197n^2-522n+840)\\
% \psi_{((()())(()))_p}=&\frac{(n-4)(n-3)^2(n-2)(n-1)}{720}(3n^2-11n+40)\\
% \psi_{((()())()())_p}=&\frac{(n-4)(n-3)(n-2)(n-1)}{2880}(5n^4-46n^3+275n^2-802n+1440)\\
% \psi_{((())(())())_p}=&\frac{(n-4)(n-3)(n-2)(n-1)}{720}(n^4-10n^3+52n^2-153n+270)\\
% \psi_{((())()(()))_p}=&\frac{(n-4)(n-3)^2(n-2)}{960}(n^4-8n^3+45n^2-78n+120)\\
% \psi_{((())()()())_p}=&\frac{(n-4)(n-3)(n-2)}{4320}(2n^6-27n^5+215n^4-993n^3+2915n^2\\&-4668n+4680)\\
% \psi_{(()((()))())_p}=&\frac{(n-4)}{2520}(10n^6-135n^5+853n^4-3378n^3+9343n^2\\&-17403n+18270)\\
% \psi_{(()(())()())_p}=&\frac{(n-4)(n-2)(n-1)}{10080}(6n^6-93n^5+771n^4-4019n^3\\&+13963n^2-30540n+35280)
%\end{align*}

\section{Proof of an antisymmetrization identity} \label{zeil}
In this section we will follow the procedure of Zeilberger \cite{Zeil-qKZ}. 
We want to prove the following equality:
\begin{equation}\label{theeq}
 \mathcal{A}\left(\frac{\prod_{1\le i<j\le r}(1+\tau u_j +u_i u_j)}{\prod_{i=1}^r u_i^{2i+p-2}(1-\prod_{j=1}^i u_j^2)}\right)_{\leq}=
\mathcal{A}\left(\prod_{i=1}^r \frac{(1+\tau u_i)^{i-1}}{u_i^{2i+p-2}\prod_{j=i}^r(1-u_j u_i)}\right)_{\leq}
\end{equation}
where $\mathcal{A}$ is the antisymmetrization operation on the variables $u_1,\ldots,u_r$, and the subscript $_\leq$ means that we only are considering the monomials of the kind $\propto \prod u_i^{a_i}$ with $a_i\leq 0$ for all $i$. 

We can directly antisymmetrize the right term:
\begin{equation}
 \mathcal{A}\left(\prod_i\frac{(1+\tau u_i)^{i-1}}{u_i^{2i+p-2}\prod_{j\geq i}(1-u_j u_i)}\right)_{\leq}=
\left(\frac{\Delta(u_j^{-1})\prod_{j>i}(u_j^{-1}+u_i^{-1}+\tau)}{\prod_i u_i^p\prod_{j\geq i}(1-u_ju_i)}\right)_{\leq}
\end{equation}
where $\Delta(u_i^{-1})=\prod_{j>i} (u_j^{-1}-u_i^{-1})$ is the Vandermonde determinant. 

In what follows
we shall call the two sides of equality \eqref{theeq} before the truncation
with $_\le$, $A_{p,r}(u_1,\ldots,u_r)$ and $B_{p,r}(u_1,\ldots,u_r)$ respectively.

The proof will be done by induction. 
The first step is to calculate the case $r=1$:
\begin{equation*}
 \mathcal{A}\left(\frac{1}{u_1^p(1-u_i^2)}\right)_{\leq}=\left(\frac{1}{u_1^p (1-u_1^2)}\right)_{\leq}
\end{equation*}

Next, we suppose that $A_{p,r-1}(u_1,\ldots,u_{r-1})_\leq=B_{p,r-1}(u_1,\ldots,u_{r-1})_\leq$.
We have
\[
 A_{p,r}(u_1,\ldots,u_r)=
\left(\sum_j(-1)^{r-j}\frac{\prod_{i\neq j}(1+\tau u_j+u_i u_j)}{u_j^{2r+p-2}(1-\prod^r_i u_i^2)}
A_{p,r-1}(u_1,\ldots,\hat{u}_j,\ldots,u_r)\right)_{\leq}
\]

Using the hypothesis and the fact that $_\leq$ is a linear operator, we rewrite the conjecture as:
\[
 B_{p,r}(u_1,\ldots,u_r)_\leq=\left((-1)^{r-j}\sum_j \frac{\prod_{i\neq j}(1+ \tau u_j +u_i u_j)}{u_j^{2r+p-2}(1-\prod_i^r u_i^2)} B_{p,r-1}(u_1,\ldots,\hat{u}_j,\ldots,u_r) \right)_\leq
\]
Working a little bit the expression, we obtain:
\[
 {B_{p,r}}_\leq=\left(\sum_j \frac{\prod_i(1-u_j u_i)}{u_j^{2r-2} (1-\prod_i u_i^2)} \prod_{i\neq j}\frac{1+\tau u_j +u_i u_j}{(u_j^{-1}+u_i^{-1}+\tau)(u_j^{-1}-u_i^{-1})}B_{p,r}\right)_\leq
\]
This equality is a consequence of
the following identity, which was pointed out in \cite{Zeil-qKZ}:
\[
\sum_j \frac{\prod_i(1-u_j u_i)}{u_j^{2r-2} (1-\prod_i u_i^2)} \prod_{i\neq j}\frac{1+\tau u_j +u_i u_j}{(u_j^{-1}+u_i^{-1}+\tau)(u_j^{-1}-u_i^{-1})}=1
\]

In order to prove this identity, we replace $u_j\rightarrow u_j^{-1}$ for all $j$:
\[
 (-1)^{r-1}\sum_j \frac{\prod_i(1-u_j u_i)}{1-\prod_i u_i^2} \prod_{i\neq j}\frac{1+\tau u_i +u_i u_j}{(u_j+u_i+\tau)(u_j-u_i)}=1
\]

Or, written in another form:
\begin{equation}
 \sum_j \frac{(\tau+2u_j)\prod_i(1-u_j u_i)(1+u_i(\tau +u_j))}{\prod_{i\neq j}(u_j-u_i)\prod_i(\tau+ u_i +u_i)}=(-1)^{r-1}(1-\prod_i u_i^2) \label{zeileq}
\end{equation}

We recall how to prove this identity using the Lagrange interpolation formula:
\begin{theo*}
Let be $P(z)$ a $N-1$ (or less) degree polynomial in $z$ and let be $(w_1,\ldots,w_N)$ different points. So these points define the polynomial that can be written by:
\[
 P(z)=\sum_j P(w_j) \prod_{i\neq j} \frac{z-w_i}{w_j-w_i}
\]
\end{theo*}

\begin{coro*}
 The maximal coefficient of $P(z)$ is:
\[
 \sum_j \frac{P(w_j)}{\prod_{i\neq j}(w_j-w_i)}
\]
\end{coro*}

Let be $\alpha$ and $\beta$ the two roots of $(1+\tau u_j + u_j^2)$. Let $P$ be 
\[
 P(z)=(\tau+2z)\prod_i (1+u_i(\tau+z))(1-u_i z)
\]

It is a polynomial of degree $(2r+1)$, so by the Lagrange interpolation formula, and using the points $(u_1,\ldots,u_r,-u_1-\tau,\ldots,-u_r-\tau,\alpha,\beta)$ to describe $P(z)$, we obtain the formula:
\begin{align*}
 2(-1)^r\prod_i u_i^2=&\sum_j\frac{P(u_j)}{(u_j-\alpha)(u_j-\beta)\prod_{i\neq j}(u_j-u_i)\prod_i(u_j+u_i+\tau)}\\
&+\sum_j\frac{P(-\tau-u_j)}{(-\tau-u_j-\alpha)(-\tau-u_j-\beta)\prod_i(-\tau-u_i-u_j)\prod_{i\neq j}(-u_j-\tau+u_i+\tau)}\\
&+\frac{P(\alpha)}{(\alpha-\beta)\prod_i(\alpha-u_i)(\alpha+u_i+\tau)}\\
&+\frac{P(\beta)}{(\beta-\alpha)\prod_i(\beta-u_i)(\beta+u_i+\tau)}
\end{align*}

Using $P(-\tau-u_j)=-P(u_j)$, $\alpha+\beta=-\tau$ and $\alpha \beta=1$, we observe that the first two terms are identical (and identical to the l.h.s. of \eqref{zeileq}), while the sum of the last two terms simplifies to $2(-1)^r$. Thus, we get \eqref{zeileq}. 
Proving this equality.

We can rewrite the main equality as a contour integral formula:
\begin{multline}
 \oint\ldots\oint\prod_i \frac{du_i}{2\pi i}\frac{\prod_{i<j}(u_j-u_i)(1+\tau u_j +u_i u_j)}{u_i^{2i+p-1}(1-\prod_{j=1}^i u_j^2)}\Omega(u_1,\ldots,u_r)=\\
=\oint\ldots\oint \prod_i \frac{du_i}{2\pi i}\frac{(1+\tau u_i)^{i-1}\prod_{j>i}(u_j-u_i)}{u_i^{2i+p-1}\prod_{j\geq i}(1-u_j u_i)}\Omega(u_1,\ldots,u_r)
\end{multline}
where $\Omega(u_1,\ldots,u_r)$ is a symmetric function in all $u_i$ without poles in the integration region around $u_i=0$. 

\section{Proof of lemma \ref{boites}}\label{skew}
In this section we intend to sketch the proof of lemma \ref{boites} about the value of $C_{a,\pi}$. 

To each $a$ and $\pi$ we associate a Young diagram $Y_a$ and $Y_\pi$, if $Y_a$ does not contain $Y_\pi$, $C_{a,\pi}=0$ as explained in section \ref{triangle}. If $Y_a=Y_\pi$ we obtain, trivially, $C_{a,\pi}=1$. The interesting case is when $Y_\pi \subsetneq Y_a$.

\begin{figure}
 \centering
 \includegraphics[scale=1.2]{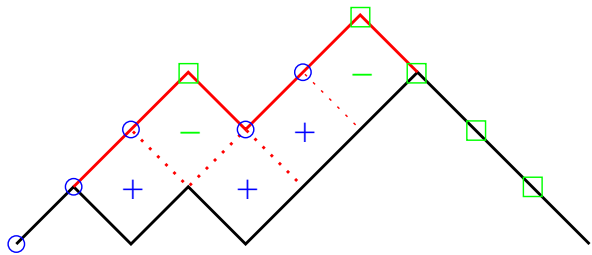}
 \caption{\label{b_Dyck}We represent the Dyck paths corresponding to $a=(1,3,5,6,7)$ (the black line, under) and $\pi=((()(())))$ (the red line, above). The openings of $\pi$ are marked with blue little circles and the closings with green little squares.}
\end{figure}

For illustration purposes we shall use an example: let $a=(1,3,5,6,7)$ and $\pi=((()(())))$, with the associated Young diagrams $Y_a=(2,2,2,1,0)$ and $Y_\pi=(1,1,0,0,0)$. We can also represent them in the form of Dyck paths as on figure \ref{b_Dyck}.

As shown in section \ref{coef}, each link contributes with a factor $U_{k}$. If a link starts at $r$ and finishes at $s$, $k$ is given by:
\[
 k=\sharp\{a_i\text{ such that }r\leq a_i <s\}-\frac{s-r+1}{2}
\]
Or, in a Dyck path representation, the $k$ is given by a counting in the NE and SE steps in the path corresponding to $a$, more precisely:
\[
 k=\frac{\sharp\{\text{NE steps between }r\text{ and }s\}-\sharp\{\text{SE steps between }r\text{ and }s\}-1}{2}
\]
which is the same as counting squares of the skew Young Diagram $Y_{a/\pi}$: those under the opening with sign plus and those under the closing with sign minus (as on figure \ref{b_Dyck}). 
In the example we get: $U_1 U_0^4=-\tau$, where the $U_1$ corresponds to the arch between positions $5$ and $8$.

If we ignore the individual arches, we note that the maximal exponent is precisely the sum all the squares (with the sign $\pm$), in the example this number is $3-2=1$.

Knowing that there is always at least one square with a minus sign, we verify that the maximal exponent of $\tau$ is $|Y_a|-|Y_\pi|-2$.

\renewcommand\MR[1]{\relax\ifhmode\unskip\spacefactor3000 \space\fi
  \MRhref{#1}{{\sc mr}}}
\renewcommand{\MRhref}[2]{%
  \href{http://www.ams.org/mathscinet-getitem?mr=#1}{#2}}

\bibliography{biblio}
\bibliographystyle{amsplainhyper}

\end{document}